\documentclass[aps,prl,twocolumn,10pt,superscriptaddress]{revtex4-2}

\usepackage{amssymb,amsfonts,amsmath}
\usepackage{times}
\usepackage{graphicx}
\usepackage{color}
\usepackage{float}
\usepackage{cancel}
\usepackage[T1]{fontenc}
\usepackage{lipsum}

\usepackage[latin9]{inputenc}
\usepackage{babel}
\usepackage{bbold}
\usepackage{wasysym}
\usepackage{multirow}
\usepackage{gensymb}
\usepackage[dvipsnames]{xcolor}
\usepackage{pifont}
\usepackage{microtype}
\usepackage{physics}
\usepackage{xcolor}

\usepackage[linktocpage=true,
  colorlinks=true, 
  pdfborder={0 0 0},
  linkcolor=blue,
  citecolor=red,
  filecolor=yellow,
  urlcolor=blue,
  bookmarks,
  pdfauthor={},
]{hyperref}

\newlength{\figurewidth}
\newlength{\pagewidth}
\usepackage{longtable}

\DeclareMathAlphabet\mathbfcal{OMS}{cmsy}{b}{n}

\begin{document}

\title{Efficient Modelling of Anharmonicity and Quantum Effects in PdCuH$_2$ with Machine Learning Potentials}

\author{Francesco Belli}\affiliation{Department of Chemistry, State University of New York at Buffalo, Buffalo, NY 14260-3000, USA}
\author{Eva Zurek}\email{ezurek@buffalo.edu}\affiliation{Department of Chemistry, State University of New York at Buffalo, Buffalo, NY 14260-3000, USA}

\begin{abstract}
Quantum nuclear effects and anharmonicity impact a wide range of functional materials and their properties. One of the most powerful techniques to model these effects is the Stochastic Self-Consistent Harmonic Approximation (SSCHA). Unfortunately, the SSCHA is extremely computationally expensive, prohibiting its routine use. We propose a protocol that pairs machine learning interatomic potentials, which can be tailored for the system at hand via active learning, with the SSCHA. Our method leverages an upscaling procedure that allows for the treatment of supercells of up to thousands of atoms with practically minimal computational effort. The protocol is applied to PdCuH$_x$ ($x = 0-2$) compounds, chosen because previous experimental studies have reported superconducting critical temperatures, $T_\text{c}$s, as high as 17~K at ambient pressures in an unknown hydrogenated PdCu phase. We identify a $P4/mmm$ PdCuH$_2$ structure, which is shown to be dynamically stable only upon the inclusion of quantum fluctuations, as being a key contributor to the measured superconductivity. For this system, our methodology is able to reduce the computational expense for the SSCHA calculations by $\sim$96\%. The proposed protocol opens the door towards the routine inclusion of quantum nuclear motion and anharmonicity in materials discovery.
\end{abstract}

\maketitle


\section{Introduction}
In the framework of standard Density Functional Theory (DFT), two fundamental approximations are commonly made. The first is concerned with the classical treatment of the nuclei, assumed to be fixed points in space generating an effective potential within which the electrons move. The second is the harmonic treatment of this potential while calculating the vibrational properties, which is approximated through a second-order expansion in the vicinity of its local minima. These approximations are key for enabling the calculation of thermodynamic properties such as the harmonic free energy, the harmonic vibrational spectra, and the heat capacity of chemical compounds and materials~\cite{baroni2001phonons}.

However, in some circumstances these fundamental approximations are no longer valid. Two examples are systems at high temperatures whose atoms undergo large amplitude atomic vibrations, or those containing light nuclei whose quantum oscillations cannot be disregarded. In both cases the atoms explore features of the potential extending far beyond the limits of a second-order expansion around the local minima~\cite{wei2021phonon}, defining the presence of anharmonicity, driven by temperature, or quantum nuclear effects. Anharmonicity is key for a broad range of materials properties including the emergence of charge density waves (CDW) in materials~\cite{gutierrez2023purely,diego2021van}, the enhancement or suppression of ferroelectricity in perovskites~\cite{fields2023temperature}, and thermodynamic properties of materials at planetary temperatures~\cite{blancas2024thermodynamics}. The effects of anharmonicity are most prominent in hydrogen-rich systems, where they can alter structural phase stability, renormalize phonon spectra and dramatically modify superconducting properties~\cite{belli2022impact,hou2021quantum,hou2021strong,errea2020quantum,errea2016quantum}.

The importance of anharmonic effects has led to the development of a number of computational techniques that can be used to treat them. One class of approaches is based upon molecular dynamics simulations carried out above the Debye temperature (assuming classical nuclei)  \cite{hellman2011lattice,hellman2013temperature2,hellman2013temperature,ljungberg2013temperature,magduau2013identification,wang1990tight,zhang2014phonon,Magdau}, or those of the path integral variant~\cite{Ceperley} (both of these methods are able to produce accurate results in the low temperature quantum limit). Another approach to describe anharmonicity to a non-perturbative level is based upon the self-consistent harmonic approximation (SCHA)~\cite{hooton1955li}, whose latest incarnation is termed the Stochastic SCHA (SSCHA)~\cite{ErreaSSCHA, Raffaello, Raffaello2,Lorenzo,Lorenzo2}. The SSCHA is able to comprehensively treat anharmonicity arising from finite temperature and from quantum nuclear effects, and it has been applied to a variety of systems including the high pressure hydrides~\cite{belli2022impact,hou2021quantum,hou2021strong,errea2020quantum,errea2016quantum}. Unfortunately, the predictive capacity of all of these methods has been limited by the computational cost associated with their requisite molecular dynamics or Monte-Carlo simulations, thereby imposing a limitation on the size of systems that can be modelled and the accuracy that can be attained. 

In addressing this challenge, Machine Learning Potentials (MLPs) emerge as a viable solution \cite{behler2016perspective, bartok2017machine, deringer2019machine, noe2020machine, westermayr2021perspective, blank1995neural}. MLPs rely on adaptable mathematical expressions used to derive analytical potential energy surfaces. The parameters of these expressions are trained on electronic structure reference data that usually includes geometric coordinates, as well as their associated energies, forces and stresses (EFS). As a result, MLPs are able to preserve the high accuracy of electronic structure methods, but at a much lower computational cost. Some of the most popular MLP variants include neural network potentials \cite{ behler2007generalized, unke2021spookynet}, Gaussian approximation potentials (GAPs) \cite{bartok2010gaussian}, and moment tensor potentials (MTPs)~\cite{shapeev2016moment,novikov2020mlip}. These potentials have proven to be powerful tools in atomistic modelling including accelerating materials discovery~\cite{merchant2023scaling,roberts2024machine}, reducing computational load, upscaling simulations to bigger sizes with minimal computational effort~\cite{leverant2020machine}, and in simulating superstructural ordering and mixed chemical occupancy \cite{CerasoliA}. Additionally, MLPs have already been used to capture anharmonicity and quantum fluctuations, setting the stage for the work described herein~\cite{fields2023temperature,lucrezi2023quantum,lam2020combining}. 

In what follows, an efficient protocol to combine MTPs with the SSCHA to model the effect of anharmonicity and quantum nuclear effects is proposed. Although MLPs have already been applied in conjunction with the SSCHA \cite{fields2023temperature,lucrezi2023quantum} there are still open questions related to the proper efficient training of the potential and its accuracy when extrapolating to larger supercells. Here these questions are addressed by creating an active learning procedure combining the SSCHA with MTPs through the MLIP (Machine Learning Interatomic Potential) package \cite{novikov2020mlip}. The outlined methodology allows for a drastic reduction of the computational cost while providing a solid framework to assess the accuracy and scalability of the developed MLPs.

Further, the computational framework is applied to PdCuH$_{2}$, a cousin of PdH -- a highly anharmonic conventional superconductor famous for its inverse isotope effect wherein the superconducting critical temperature,  $T_\text{c}$, of the deuterated form is higher than that of its lighter brethren \cite{skoskiewicz1973superconductivity}. In 1974 a PdCuH$_x$ compound with $x \sim1.4$ was synthesized and its $T_\text{c}$ was reported to be 17~K~\cite{stritzker1974high}, nearly double the value measured for PdH. A recent computational study concluded that the most likely candidate for the superconducting phase was a $P4/mmm$ PdCuH$_2$ stoichiometry structure, but an accurate calculation of its $T_\text{c}$ could not be performed due to the presence of imaginary phonon modes, which were hypothesized to be a manifestation of the strong anharmonicity~\cite{vocaturo2022prediction}. In what follows, we  show that quantum nuclear effects render this PdCuH$_2$ phase dynamically stable, yielding a $T_\text{c}$ that is in-line with experiment. Additionally, we searched for a possible inverse isotope effect by calculating the $T_\text{c}$ of PdCuD$_2$, but none was found.

\section{Results}
\subsection*{SSCHA + MLIP Workflow}
The first step of the SSCHA involves the generation of a set of displaced atomic configurations, sampled from a Gaussian distribution with a width linked to the magnitude of the computed force constants~\cite{Lorenzo}. Consequently, an initial guess for the dynamical matrices, typically obtained within the harmonic approximation for the chosen supercell, is required before initialising the SSCHA. For accurate results, the chosen supercell should be large enough to capture the necessary atomic interactions. In the procedure implemented here the initial guess is made for a supercell smaller then the one needed for convergence -- foreshadowing the upscaling procedure that distinguishes this workflow from those proposed by others.

\begin{figure*}[th!]
\centering
\includegraphics[width=0.8\textwidth]{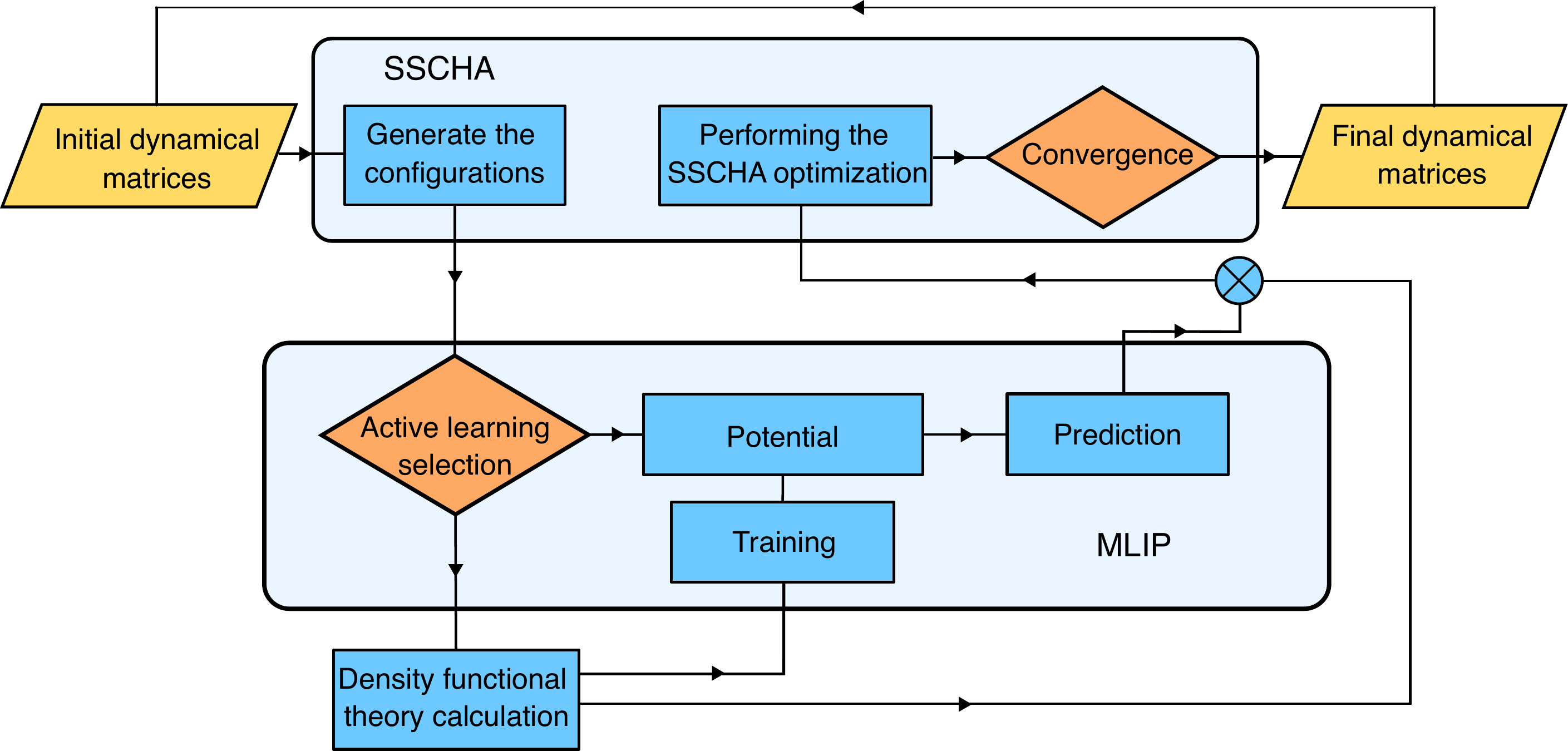}
\caption{\label{Fig:SSCHA_MLIP_Workflow} \textbf{SSCHA + MLIP workflow.} Flowchart for the workflow that combines the Stochastic Self-Consistent Harmonic Approximation (SSCHA) with the Machine Learning Interatomic Potential (MLIP) package for the generation of moment tensor potentials (MTPs). In this flowchart, the blue rectangles denote individual workflow steps, the orange rhombi indicate logic operations, while the generic yellow parallelograms represent input and output files. Large rounded rectangles encompass sections of the program related to the SSCHA and MLIP packages. This workflow fully integrates both programs and incorporates an on-the-fly procedure for MTP training, significantly reducing the number of required DFT calculations and automating dataset training.}
\end{figure*}
%
It is at this step that the SSCHA+MLIP integration, based upon an active learning protocol, begins: the workflow is depicted schematically in Figure \ref{Fig:SSCHA_MLIP_Workflow}. Displaced configurations for the supercell are created, and single point DFT calculations are performed to obtain the EFS of all of the members of this first population, to be used for training the MTP. In subsequent populations, the extrapolation grade, $\gamma$, of each structure is determined via the generalised D-optimality criterion, which is able to detect the most diverse configurations ~\cite{novikov2020mlip}. To ensure accurate predictions we use a very conservative value of $\gamma_\text{select}$ = 2, chosen so that every geometry with $\gamma_\text{select} >$  2 has their structures and EFS added to the training set. This provides a strict threshold for the selection of the training set.

The retrained MTP is then employed to predict the EFS of the members of this population not chosen for training.  After the EFS of all of the configurations in the population have been obtained (via MTP or DFT) they are combined and fed to the SSCHA to perform the optimisation of the force constants and centroid positions. This procedure is repeated for subsequent populations until the SSCHA is converged. Though the MTP parameters are updated for each population, they are not used to predict the EFS of members of prior populations because of the cautious value of $\gamma_\text{select}$ and because in the SSCHA each geometry is independent of the others, unlike in a molecular dynamics run or in a relaxation trajectory where the same set of MTP parameters must be employed to avoid the introduction of discontinuities in the potential energy surface. 
\begin{figure*}[t]
\centering
\includegraphics[width=\textwidth]{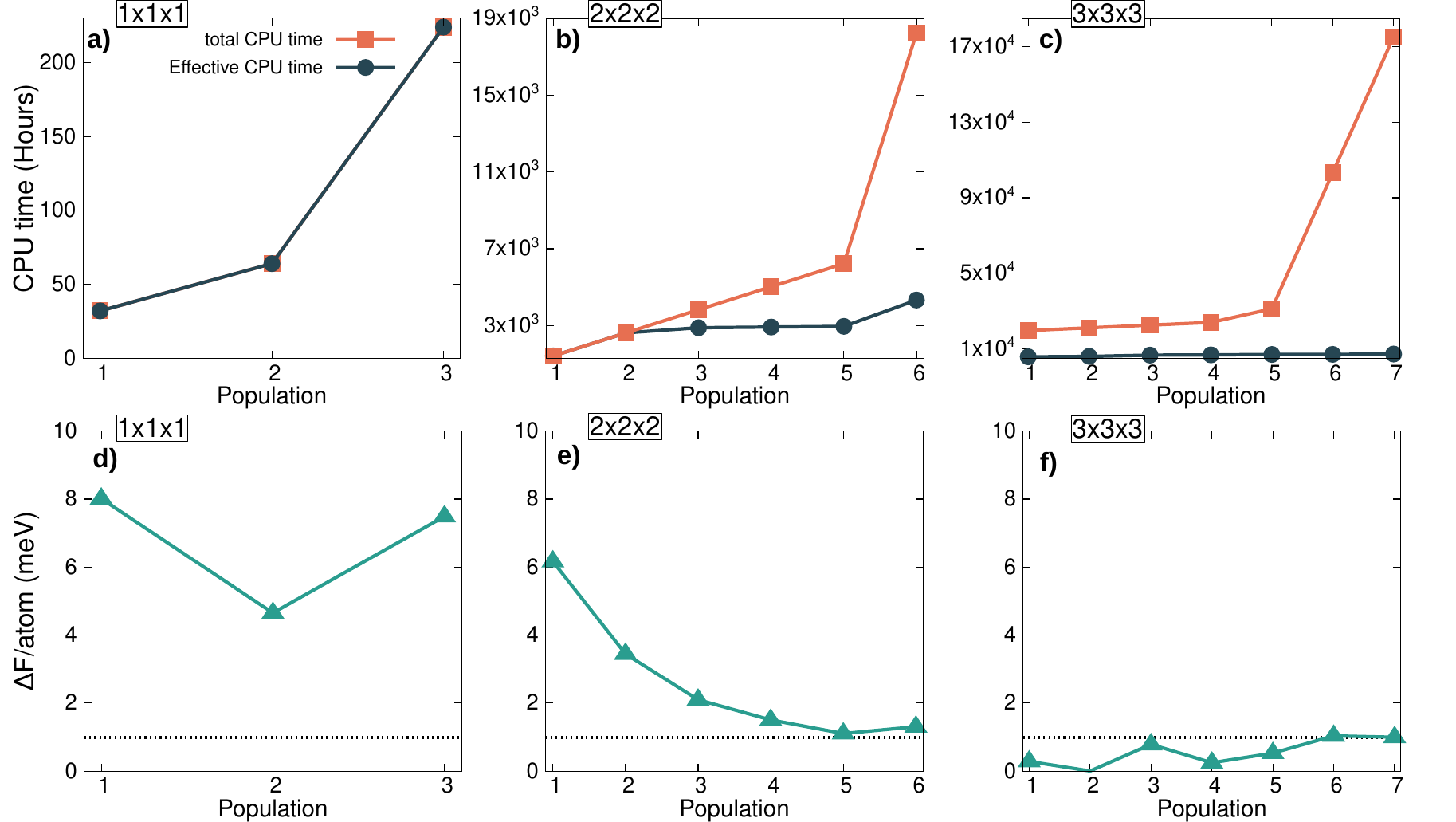}
\caption{\label{Fig:Scaling} \textbf{Effective process time.}  The CPU time, in hours, required for executing the full active learning and upscaling procedure starting with the (a) 1$\times$1$\times$1 cell and extending to the (b) 2$\times$2$\times$2 and the (c) 3$\times$3$\times$3 supercells. The orange lines represent the total CPU time that would be necessary without the machine learning protocol; the dark green lines represent the actual CPU time utilised. In (a) these lines overlap. The evolution of the difference between the anharmonic free energy, $\Delta F$, relative to its minimum calculated value as a function of the population during training for (d) 1$\times$1$\times$, (e) 2$\times$2$\times$2 and (f) 3$\times$3$\times$3 supercells. The potential was trained at a temperature of 300~K and can be used to reproduce supercells larger than 3$\times$3$\times$3.}
\end{figure*}

The cycle consisting of generating displaced atomic configurations, passing them through the active learning procedure for EFS predictions, and the SSCHA optimization is repeated until the SSCHA reaches convergence. Subsequently, an upscaling procedure is performed, in which the SSCHA is reinitialized with an increased supercell size. This step is crucial for properly modelling phenomena such as CDWs, which might not be captured in smaller cells. A tight-binding model is employed to extrapolate the starting guess for the new upscaled dynamical matrices, from the converged dynamical matrix obtained in the previous smaller cell. 

The SSCHA+MLIP cycle is repeated again, except now the first iteration of the SSCHA employs the final MTP developed from the previous run on the smaller cell, instead of DFT. As before, when extrapolative configurations are encountered they are stored and may be used to retrain the MTP. This upscaling process is then iterated until reaching the desired size for the supercell. The rationale underlying the upscaling procedure is rooted in the idea that the majority of atomic interactions required for training the potential are already present in the small cells \cite{meziere2023accelerating} and that progressively fewer configurations are required to retrain the potential when considering bigger cells. 

The results of the training procedure for the specific case of $P4/mmm$ PdCuH$_{2}$, whose phonon spectra and superconducting properties are investigated in the following section, is reported in Figure \ref{Fig:Scaling}. The SSCHA procedure began from calculations performed on the four atom unit cell. In the first three populations all of the 140 configurations were evaluated by DFT. Consequently, the yellow points (CPU time required for a traditional SSCHA run) overlap with the green points (CPU time required for the MLIP accelerated SSCHA) in the top panel. After these three populations, the SSCHA was converged for the unit cell and the calculations were upscaled to a 2$\times$2$\times$2 supercell containing 32 atoms. In this supercell, the SSCHA was converged after the 6$^\text{th}$ population and only 342 out of the 1500 configurations (22.8\%) were deemed extrapolative requiring further DFT calculations, whose results were employed to retrain the potential. The system was further upscaled to a 3$\times$3$\times$3 supercell containing 108 atoms; from the 2180 configurations required for SSCHA convergence only 56 (2.57\%) were sent for DFT calculations. After the training on the 3$\times$3$\times$3 supercell was complete, the MTP did not require any additional DFT data while upscaling to even larger cells. Though the potential was explicitly trained on a cells containing up to 108 atoms, it could be used to perform the SSCHA on a 6$\times$6$\times$5 supercell containing 720 atoms. In this specific case 12,000 configurations and four populations were necessary to achieve convergence for the 6$\times$6$\times$5 supercell. Because of the formal $N^3$ scaling in DFT and the increasingly large number of configurations required by the SSCHA for larger unit cells, such calculations would have been impossible without the MLP acceleration. When taking into account the amount of time required to converge the 3$\times$3$\times$3 supercell, the SSCHA + MLIP workflow results in a reduction in computational resources by an impressive $\sim$96\%. A full DFT SSCHA run was not performed for the 6$\times$6$\times$5 supercell, since this would have been impossible during a typical postdoc lifetime, and particular effort was invested in safeguarding the sanity and mental well-being of the one involved in this investigation.

The accuracy and the predictive capabilities of the potential were analysed by inspecting the root mean square (RMS) prediction errors for the EFS as a function of the size of the supercell (Supplementary Figure 1). The prediction errors were obtained by comparing the DFT EFS of an auxilliary SSCHA-generated set of structures for the given supercell size with those predicted by the MTP trained on a particular supercell with the SSCHA+MLIP procedure, as described more in depth in Supplementary Note 1. Although the MTP was relatively accurate after training on a 2$\times$2$\times$2 supercell, the accuracy we desire was only achieved after training on the 3$\times$3$\times$3. The resulting MTP gives an RMS error of less than: 0.5~meV/atom for the energies, 0.05~meV/(atom~\AA\ ) for the forces, and  0.5~GPa for the stresses.  The necessity for training on the 3$\times$3$\times$3 supercell can also be noticed by inspecting the behavior of the free energy at 300~K as a function of the supercell and population, reported in Figure \ref{Fig:Scaling}. Although, one might suspect convergence is reached at the end of the 2$\times$2$\times$2 run, a small decrease in the free energy is observed when the training is completed on the 3$\times$3$\times$3 cell.

The SSCHA + MLIP workflow, including the active learning protocol and the upscaling procedure, is more efficient and complete than recently proposed workflows~\cite{fields2023temperature,lucrezi2023quantum}. A key advantage of our approach is that it enables the automation of the training and facilitates the evaluation of its efficacy, ensuring accuracy in the predictions, via the active learning phase.  The previously established methods require prior knowledge of the final structure and its environment to effectively train the potential. In contrast, the active learning protocol proposed here enables the use of an arbitrary initial structure. Additionally, the accuracy of this model was evaluated by comparing its errors to those reported by Lucrezi et al.~\cite{lucrezi2023quantum}. For a 2$\times$2$\times$2 supercell and a level 10 potential, our MTP errors are 0.5-0.6~meV/atom for the energies, 20-30~meV/{\AA} for the force components, and 0.2-0.3~GPa for the diagonal stress tensor components. These results are in good agreement with those presented by Lucrezi et al.~\cite{lucrezi2023quantum}. More details on the errors obtained for the potential training are reported in Supplementary Note 1.

Additionally, this methodology not only provides the possibility of drastically decreasing the computational cost required for calculating the EFS during the upscaling procedure, but it also enables the identification of emerging phenomena such as CDWs, which potentially require retraining of the potential. Furthermore, a MLP developed at a particular temperature and pressure can be employed as a starting guess for a simulation performed for a different set of thermodynamic variables. For example, an MTP trained at high temperatures could be used for a calculation performed at low temperatures, requiring only a few additional DFT calculations for retraining via the active learning protocol. Similarly, an MTP trained at a few pressures in a narrow pressure range could be employed to perform the SSCHA dynamics allowing for a full cell relaxation with minimal computational expense.

\begin{figure}[t]
\centering
\includegraphics[width=\columnwidth]{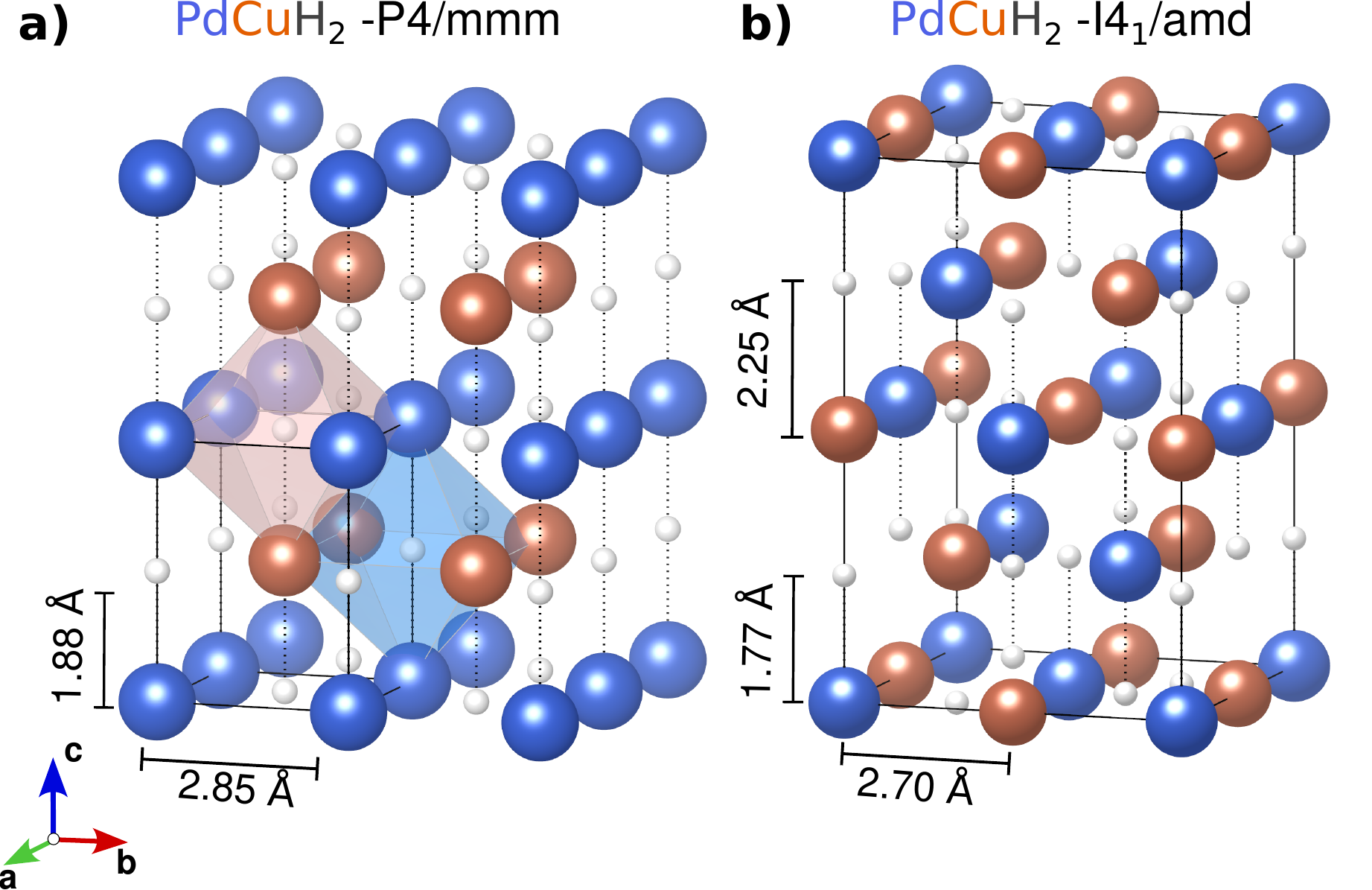}
\caption{\textbf{Crystal structures.} The most stable PdCuH$_2$ phases proposed in (a) Ref.\ \cite{vocaturo2022prediction}, and (b) that was found here. Both studies identified the same PdCuH phase as the ground state. It can be obtained from $P4/mmm$ PdCuH$_2$ by removing the hydrogen atoms that are at the same sites as the hydrogen in the center of the blue transparent octahedron.}
\label{fig:struc}
\end{figure}

\subsection*{PdCuH$_x$ ($\mathbf{x=0-2}$) Structures through Crystal Structure Prediction} 
As the hydrogen content and positions within the synthesized PdCuH$_x$ compound are unknown~\cite{stritzker1974high}, crystal structure prediction (CSP) calculations with the \textsc{XtalOpt} evolutionary algorithm were performed for $x=0-2$ at zero temperature and pressure with 1-4  formula units in the cell. The equimolar PdCu alloy adopts a disordered fcc structure and an ordered CsCl (B2) phase, and at high temperatures experiments find a mixture of the two~\cite{subramanian1991cu,zhang2017palladium}. Our CSP searches identified the B2 ($Pm\bar{3}m$ symmetry) ordered phase as being the ground state at 0~K, verifying the CSP framework and choice of computational parameters. 

The small hydrogen atoms are often adsorbed into the interstitial sites of transition metal solids. Therefore, it was not a surprise that the most stable PdCuH phase found in our CSP runs could be obtained from $Pm\bar{3}m$  PdCu by filling half of its octahedral sites, specifically those that are coordinated equatorially by four palladium atoms and axially by two copper atoms (\textit{cf.}\ Figure \ref{fig:struc} (a)) with hydrogen. This structure was also predicted to be the ground state by Vocaturo \textit{et al.}\ who found it via random placement of hydrogen within the parent PdCu unit cell followed by local relaxation~\cite{vocaturo2022prediction}.  Filling the second octahedral site, this time with copper on the equatorial and palladium on the axial positions leads to the most stable PdCuH$_2$ structure identified by Vocaturo and co-workers with $P4/mmm$ symmetry~\cite{vocaturo2022prediction}, as shown in Figure \ref{fig:struc} (a). Another PdCuH$_2$ phase they identified, which was $\sim$50~meV/atom higher in energy than $P4/mmm$  PdCuH$_2$, was characterized by hydrogen atoms on both of the tetrahedral holes, highlighting the preference of hydrogen occupying the octahedral rather than the tetrahedral sites in hydrogenated PdCu alloys.

The evolutionary searches performed here, on the other hand, predicted an $I4_1/amd$ symmetry PdCuH$_2$ structure (Figure \ref{fig:struc}(b)) as being the ground state. It was calculated to be 18~meV/atom \emph{more stable} than the previously proposed $P4/mmm$ PdCuH$_2$. This newfound $I4_1/amd$ symmetry phase is also characterized by hydrogen-filling of the octahedral sites, however it differs from $P4/mmm$ in the arrangement of the metal atoms. Whereas the latter contains alternating layers of Pd and Cu stacked in an ABA... fashion, in $I4_1/amd$ each layer is constituted by an equal proportion of Cu and Pd atoms with an ABCDA... stacking along the $c$-axis. In $I4_1/amd$ each hydrogen atom occupies the center of an octahedron where two copper atoms and two palladium atoms, diagonal to each other, constitute the equatorial sites. Similarly, one of each type of metal atom lies on the axial positions. As a result, the Pd-H and Cu-H distances are the same in the equatorial plane, but they differ along the $c$ lattice vector with slightly shorter Pd-H (1.97~{\AA}) and slightly longer Cu-H (2.25~{\AA}) distances so that the hydrogen atoms do not all lie in the same $ab$ plane, unlike in $P4/mmm$ PdCuH$_2$.

A final relaxation of the crystal structure was performed through the SSCHA at zero temperature for both the $P4/mmm$ and the $I4_1/amd$ phases after performing CSP. In this relaxation the unit cell shape and volume were held fixed, while allowing the atomic positions to relax, subject to symmetry constraints. No significant modifications to the crystal structure were observed as a result of this relaxation. Furthermore, the introduction of quantum nuclear effects did not induce any anisotropic pressure contributions, eliminating the necessity for a further relaxation of the unit cell.

\subsection*{Electronic Structure and Superconductivity} 
Now that we have described the structural peculiarities of the most stable PdCuH phase identified via both random~\cite{vocaturo2022prediction} and evolutionary searches, as well as two PdCuH$_2$ phases -- a metastable one that can be obtained by filling the octahedral holes of B2 PdCu with hydrogen, and the ground state, which possesses a different arrangement of the transition metal atoms, let us investigate their electronic structure and propensity for superconductivity. Specifically, we aim to determine which of these compounds, if any, could be a potential contributor to the measured $T_\text{c}$ of 17~K~\cite{stritzker1974high} in a hydrogenated PdCu sample. Since previous computations classified PdCuH as a sub-Kelvin superconductor~\cite{vocaturo2022prediction}, we turned our focus to the PdCuH$_2$ stoichiometry.

\begin{figure}[h!]
\centering
\includegraphics[scale=0.38]{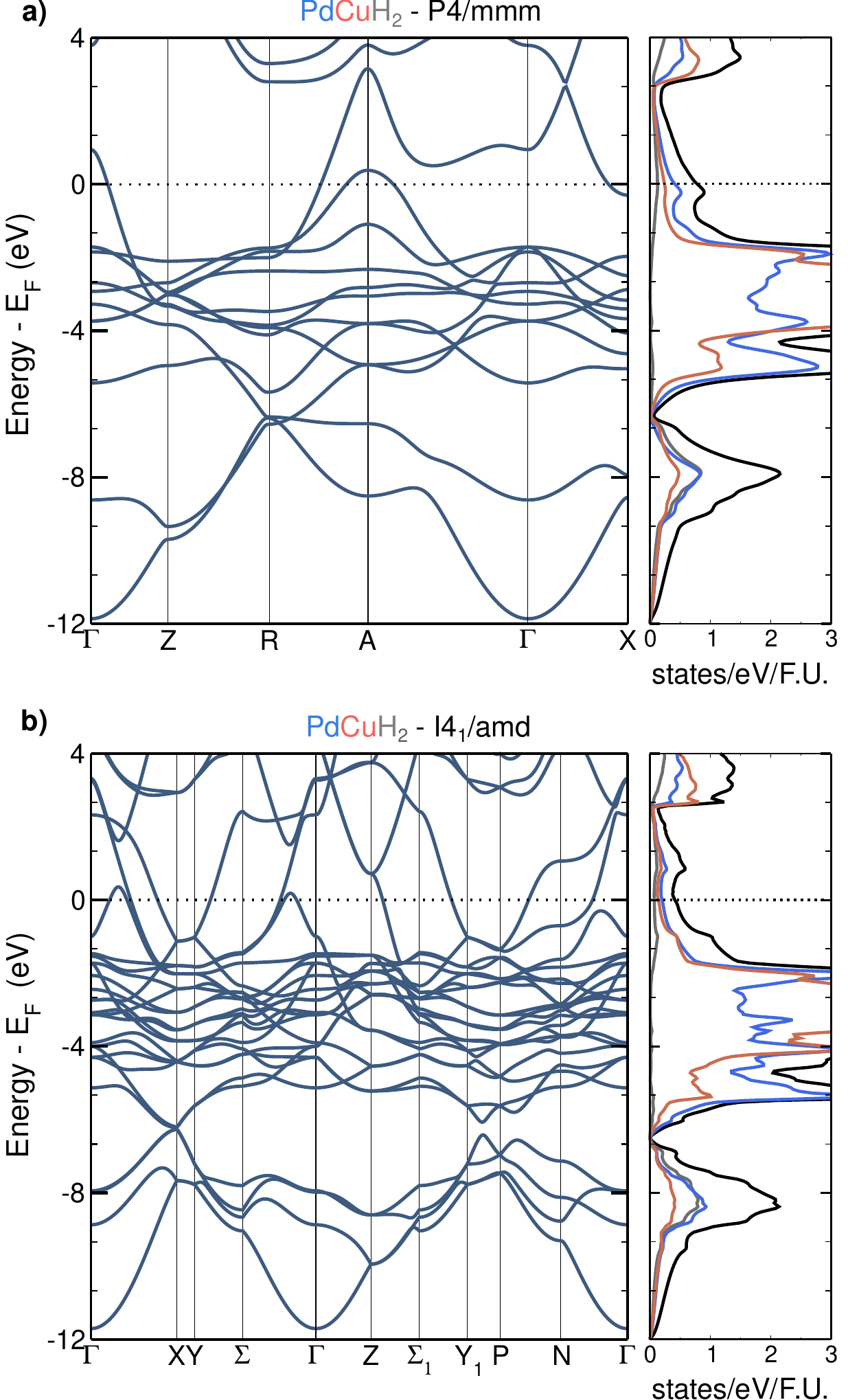}
\caption{\label{Fig:E_band_structure} \textbf{Electronic band structures.} Electronic band structure and atom projected densities of states (DOS) of (a) $P4/mmm$ PdCuH$_2$ and (b) $I4_1/amd$ PdCuH$_2$. The total DOS is given by the black line.}
\end{figure}

Figure \ref{Fig:E_band_structure} reports the electronic band structure and the density of states (DOS) for both $P4/mmm$ and $I4_1/amd$ PdCuH$_2$. Whereas the former contains one formula unit (FU) in the primitive cell, the unit cell of the latter is doubled, and therefore so is the total number of bands. Importantly, the DOS at the Fermi level, $E_\text{F}$, in the lower-energy $I4_1/amd$ structure is only half the value as in the $P4/mmm$ phase, 0.424 vs.\ 0.784 states eV$^{-1}$ FU$^{-1}$, respectively. A computational experiment where the DOS was calculated for two hypothetical PdCuH$_2$ structures: (i) with the same PdCu lattice as in the $I4_1/amd$ phase but with the hydrogen atoms situated in the center of the octahedral holes, and (ii) with the same distorted hydrogen lattice as in the $I4_1/amd$ phase, but with the metal atoms arranged as in the $P4/mmm$ geometry, revealed that the metal atom placement was the main driver of the decreased DOS in the ground state.

In both structures the contributions to the DOS at $E_\text{F}$ were the lowest for hydrogen, followed by copper then palladium (16\%, 31\% and 53\% for $P4/mmm$ vs.\ 16\%, 35\% and 49\% for $I4_1/amd$). The profile of the DOS for both PdCuH$_2$ structures resembles the one calculated for pure PdH~\cite{yang2017formation}. However, the extra half electron per MH unit in PdCuH$_2$  shifts the Fermi energy further away from the bulk of the d states of both copper and palladium, with a concomitant reduction of the DOS at $E_\text{F}$ in the ternary hydride by almost a factor of two. 

\begin{figure*}[t]
\centering
\includegraphics[width=\textwidth]{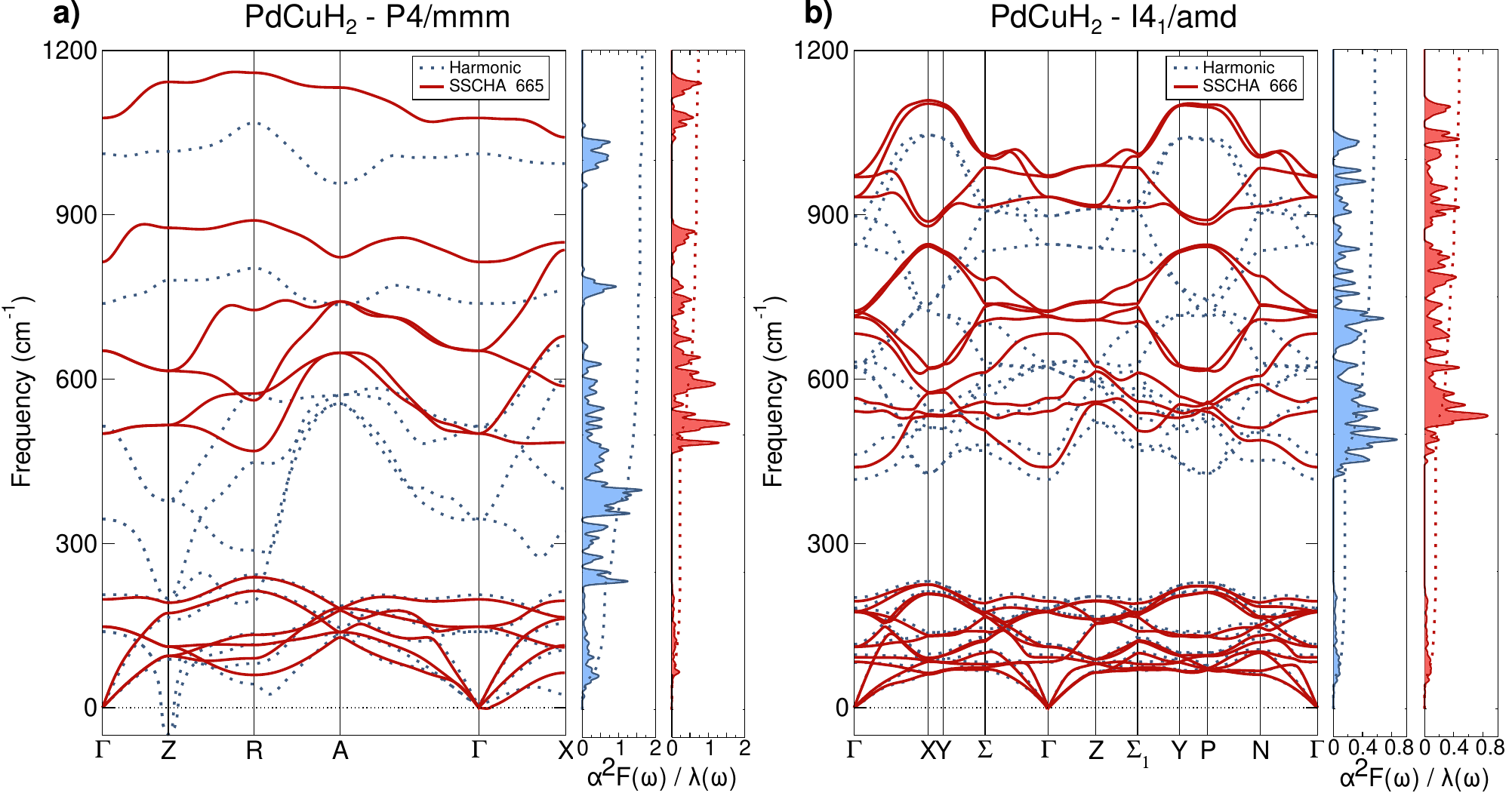}
\caption{\label{Fig:Phonons}\textbf{Phonon band structures.} The phonon band structure, Eliashberg spectral function, $\alpha^2F(\omega)$, and the integral of the electron phonon coupling, $\lambda(\omega)$,  for (a) $P4/mmm$ PdCuH$_2$ and (b) $I4_1/amd$ PdCuH$_2$ calculated with classical nuclei (blue dashed lines and shading) and incorporating the effects of  anharmonicity and quantum nuclei (red solid lines and shading).}
\end{figure*}

The phonon spectra for both PdCuH$_2$ phases is shown in Figure \ref{Fig:Phonons}. Since the $P4/mmm$ unit cell contains four atoms while eight atoms comprise the $I4_1/amd$ unit cell, a total of 12 and 24 phonon modes, respectively, is obtained. In both phases, the phonons can be divided into three distinct branches: the six (or 12) lowest frequency modes are related to the acoustic vibrations of both the Pd and Cu atoms, while the intermediate and upper branches, having two (or four) and four (or eight) modes respectively, correspond to optical vibrations linked to the  hydrogen atoms. 

The classical $P4/mmm$ PdCuH$_2$ phonons (blue dashed curve) reveal an instability at the $Z$ high-symmetry point and a softening at $X$, and they are in good agreement with those previously calculated by Vocaturo and co-workers~\cite{vocaturo2022prediction}. Additionally, the intermediate optical branch associated with the hydrogen atom motions shows signs of strong anharmonicity through soft modes at $Z$ and $R$. The introduction of quantum nuclear effects (red solid line) has a profound effect on the phonon band structure. Most notably, they are  able to heal the instability at $Z$ and the soft mode at $X$ associated with the acoustic branches. Both the intermediate and upper optical branches are subject to a strong red shift, which has a greater impact on the intermediate branch that is pushed up by about 400~cm$^{-1}$. 

To ensure that the anharmonic corrections do not derive from artifacts of the local properties of the potential we used the meta-GGA R$^2$SCAN functional, as implemented in the VASP code, to calculate both the harmonic and quantum anharmonic phonons (Supplementary Figures 6 and 7). Mirroring the results obtained with the PBE functional and the Quantum Espresso code, the SSCHA healed an imaginary mode at the $Z$-point, as well as soft modes around the $R$ and $X$ points, rendering this phase dynamically stable. Furthermore, the comparison between the quantum anharmonic phonons obtained with PBE and R2SCAN (Supplementary Figure 6) suggests that the R2SCAN introduces just minor corrections to the spectra, confirming that the PBE provides a good approximation for the system.

The phonon band structure for the lower symmetry $I4_1/amd$ PdCuH$_2$ phase is markedly different. Importantly, no instabilities are present in the harmonic phonons and the introduction of quantum nuclear effects does not significantly affect the vibrational frequencies associated with a particular mode, suggesting the motions of atoms in this phase are mostly harmonic. The most notable difference between the harmonic and the SSCHA-computed phonon modes are the minor re-normalisation of the intermediate and upper optical branches, which undergo a blue shift separating them in energy. 

Finally, the $T_\text{c}$ of these two phases was estimated via the Allen-Dynes equation~\cite{allen1975transition} (Table \ref{table:tc}). For classical nuclei,  and ignoring the instability around the $Z$-point, the $T_\text{c}$ of $P4/mmm$ PdCuH$_2$ was estimated to be 37-44~K, in good agreement with previous estimates of 40-45~K, also obtained by neglecting the imaginary phonon modes~\cite{vocaturo2022prediction}. The large value of $\lambda$ is related to the soft hydrogen optical modes that enhance the Eliashberg spectral function at low frequencies, as seen in Figure \ref{Fig:Phonons}(a). 

When quantum nuclear effects are introduced, the integral of the spectral function up to $\sim$260~cm$^{-1}$ significantly decreases, and the region between 260-460~cm$^{-1}$ falls to zero, shifting upwards in frequency, with a concomitant decrease of $\lambda$ and increase of $\omega_{\textrm{log}}$. The resulting $T_\text{c}$, which incorporates the quantum nuclear fluctuations, is significantly reduced to 13-20~K, in-line with the maximum value of 17~K experimentally reported for a presumed Pd$_{1.1}$Cu$_{0.9}$H$_x$, $x\sim1.4$ phase~\cite{vocaturo2022prediction}.  About 75\% of $\lambda$ derives from the part of the Eliashberg spectral function associated with the motions of the hydrogen atoms, suggesting that they provide the key contribution to superconductivity in this phase, similar to previous results obtained for PdH~\cite{errea2013first}. However, in $P4/mmm$ PdCuH$_2$ the acoustic branches furnish a greater contribution towards $\lambda$ increasing its value by $\sim$33\% compared to PdH~\cite{errea2013first}, while $\omega_{\textrm{log}}$ remains almost unchanged with just a small reduction of 8\%. This suggests that the more-than-doubled quantum $T_\text{c}$ of $P4/mmm$ PdCuH$_2$ as compared to PdH is likely due to the reduction in mass related to the substitution of half of the palladium atoms with the lighter copper atoms. 

Considering the similarity of PdCuH$_2$ with PdH, the possible presence of the inverse isotope effect was investigated. Towards this end, the phonon spectrum, Eliashberg spectral function (Supplementary Figure 4) and the resulting $T_\text{c}$ of the deuterated compound was calculated with and without the SSCHA.  While the $\lambda$ of $P4/mmm$ PdCuH$_2$ and PdCuD$_2$ were nearly identical, the latter had a reduction of $\sim$15\% in $\omega_{\textrm{log}}$, as expected, resulting in a decreased $T_\text{c}$ of 11-17~K. Therefore, no inverse isotopic effect was found. 

The computed trends in the superconducting properties differ markedly for $I4_1/amd$ PdCuH$_2$, whose phonon band structure and Eliashberg spectral function was not strongly perturbed by the introduction of quantum nuclear effects. Assuming classical nuclei yields a $T_\text{c}$ of 5-10~K, whose value is reduced slightly to 2-5~K via the introduction of quantum nuclear effects. As compared to the $P4/mmm$ phase, the reduction, which results from the slight blue shift of the optical phonon branches that moves  $\alpha^2F(\omega)$ to higher frequencies (Figure \ref{Fig:Phonons}(b)), is not significant. The lower $T_\text{c}$ of this phase as compared to $P4/mmm$ PdCuH$_2$ is likely due to the decreased value of the DOS at the Fermi energy. 

The renormalization of the $T_\mathrm{c}$ in both structures can be predominantly linked to the change of the phonon spectra. Lucrezi et al. \cite{lucrezi2023quantum} proposed that the discrepancy in the phonon spectra between the SSCHA and the harmonic calculations can be separated into two contributions: structural renormalizations induced by quantum nuclear effects and phonon-phonon interactions driven by anharmonicity. For the PdCuH$_2$ structures, since no structural renormalization arises from SSCHA, the phonon renormalization is solely a consequence of anharmonicity-induced phonon-phonon interactions.

\begin{table}[]
\caption{\label{Table:Tc} \textbf{Superconductivity.} The values of the electron phonon coupling parameter, $\lambda$, the logarithmic average frequency, $\omega_\text{log}$ (cm$^{-1}$), and the superconducting critical temperature, $T_\text{c}$ (K), for the specified systems calculated for classical (harmonic) and quantum (SSCHA) protons.  These values can be compared with experimental reports of a maximum $T_\text{c}=$~17~K in a PdCuH$_x$ compound whose stoichiometry and structure were not determined~\cite{stritzker1974high}.}
\begin{tabular}{c|ccccccc}
\hline
\hline
                      & \multicolumn{3}{c}{Harmonic$^a$}      &&        \multicolumn{3}{c}{SSCHA}                   \\
                      & $\lambda$ & $\omega_\text{log}$ & $T_\text{c}^b$ && $\lambda$ & $\omega_\text{log}$ & $T_\text{c}^b$ \\
\hline
$P4/mmm$ PdCuH$_2$    & 1.64      & 208                 & 37-44          && 0.73      & 344                 & 13-20          \\
$I4_1/amd$ PdCuH$_2$  & 0.57      & 351                 & 5-10           && 0.48      & 349                 & 2-5            \\    
PdH$^c$               & 1.55      & 205                 & 34             && 0.40      & 405                 & 5              \\
$P4/mmm$ PdCuD$_2$    & 1.65      & 163                 & 29-34          && 0.74      & 291                 & 11-17          \\
\hline
\hline 
\end{tabular} \\
$^a$ The imaginary frequencies in the harmonic phonons of $P4/mmm$ PdCuH$_2$ and PdCuD$_2$ were neglected. \\
$^b$ $T_\text{c}$ calculated using the Allen-Dynes modified McMillan equation with $\mu^*=$~0.15 and 0.1. \\
$^c$ Reference \cite{errea2013first}.
\label{table:tc}
\end{table}

In the experiments PdCu alloys were made by arc-melting in ultrapure argon, but their atomistic arrangement was not reported~\cite{stritzker1974high}. Hydrogen was added to the resulting PdCu metal foils by precharging under an H$_2$ gas pressure of 4~atm at 300$^\circ$C followed by H$_2^+$-ion implantation. It is therefore likely that the hydrogen atoms were incorporated into octahedral holes already present in the PdCu alloys, and the position of the metal atoms did not rearrange during the hydrogenation. 

The PdCu lattice of the $I4_1/amd$ structure is 11~meV/atom less stable than the CSP-found lowest energy B2 phase, and also differs markedly from the disordered fcc structure that can be found in the alloy. Therefore, it is very unlikely that the hydrogenated alloy made in experiment contains regions that, on a local scale, resembled $I4_1/amd$ PdCuH$_2$. Thus, though this phase is thermodynamically preferred, it is kinetically disfavored, explaining why the measured maximum $T_\text{c}$ was significantly higher than the one estimated for this phase. Similarly, the good agreement between the experiments and the calculated $T_\text{c}$ of the $P4/mmm$ phase, which could be derived from the B2 PdCu phase by placing hydrogen in its octahedral holes, suggests that local regions with this motif were present in the superconducting phase, whose stoichiometry was estimated as being Pd$_{1.1}$Cu$_{0.9}$H$_{1.4}$. 

Calculations on supercell models derived from $P4/mmm$ PdCuH$_2$ where some of the hydrogen atoms are removed from the octahedral positions resulting in a $\sim$Pd$_{1.1}$Cu$_{0.9}$H$_{1.4}$ composition are directions for future work, as are studies of metastable PdCu lattices that are subsequently hydrogenated. The results presented here reveal a  $\sim$4-fold enhancement of $T_\text{c}$, depending upon the atomistic arrangement of the Pd and Cu atoms in the alloy, which in turn affect the placement of the H atoms thereby influencing the DOS at $E_\text{F}$, the vibrational properties and the electron phonon coupling. It is conceivable that $T_\text{c}$ could be raised further, and the atomistic arrangement of the alloy be engineered via synthesis routes far from equilibrium. In addition, these avenues may prove fruitful for designing hydrogenated AgPd or AuPd alloys with superconducting properties surpassing those that have been previously reported in this class of compounds~\cite{stritzker1974high}.

\section{Discussion}

This study outlines an effective protocol that combines the Stochastic Self-Consistent Harmonic Approximation (SSCHA) with the  Machine Learning Interatomic Potential (MLIP) package, enabling the automated and rapid calculation of anharmonic and quantum nuclear effects for a wide range of materials properties. This procedure enhances the performance of the SSCHA,  improving the prospect for a routine treatment of quantum nuclear and anharmonic effects in computations. In the proposed workflow DFT calculations are performed on geometries (generated by the Monte Carlo algorithm of the SSCHA) that are deemed to be extrapolative, and a subset of these is employed to train quantum-accurate moment tensor potentials (MTPs) that are used to predict the energies, forces and stresses of the remaining geometries required for the SSCHA.  Furthermore, we introduce an upscaling protocol, where the MTPs are first trained on small cells, and retrained on larger ones, allowing for the simulation of systems with sizes that were previously out-of-computational-reach. The upscaling protocol is based on the assumption that most of the interatomic interactions are already captured in the small cells, and that only minor corrections to the MTPs are required when larger cells are considered. 

Although general in nature, this procedure is not directly applicable to other types of MLPs due to the specific implementation of the active learning protocol in the MLIP program package. An extension of the procedure to GAPs, could be possible by adopting an active learning strategy similar to the one proposed in Refs. ~\cite{jinnouchi2019fly,sivaraman2020machine}, where new structures for the training set are selected on the estimation of the Bayesian error during the simulation. Alternatively, for neural networks, the active learning procedure could involve randomly selecting structures to create a validation set at each step of the simulation, which would then be used to assess the accuracy of the potential.

The protocol is put to the test on the PdCuH$_{2}$ system, which was hypothesized~\cite{vocaturo2022prediction} to be an important contributor to the superconductivity reported for hydrogenated PdCu alloys with a maximum superconducting critical temperature, $T_\text{c}$, of 17~K at ambient pressure~\cite{stritzker1974high}. Hydrogenation of the most stable $Pm\bar{3}m$ symmetry PdCu phase would likely yield the kinetically favored $P4/mmm$ PdCuH$_2$ structure, metastable by 18~meV/atom from the newly predicted $I4_1/amd$ PdCuH$_2$ ground state. Similar to the highly anharmonic PdH phase, $P4/mmm$ PdCuH$_2$ is found to be dynamically stable only upon the inclusion of quantum and anharmonic effects, which heal the imaginary modes in its phonon band structure, and renormalize the frequencies of others. These effects decrease the predicted $T_\text{c}$ of $P4/mmm$ PdCuH$_2$ from 37-44~K to 13-20~K, mirroring previous trends computed for PdH~\cite{errea2013first}, whose anharmonic $T_\text{c}$ is notably smaller ($\sim$5~K), likely because of the lighter mass of the copper, as compared to that of palladium. An inverse isotope effect on superconductivity was not observed when hydrogen was replaced by deuterium, with an estimated $T_\text{c}$ of 11-17~K for $P4/mmm$ PdCuD$_2$. The thermodynamically preferred $I4_1/amd$ PdCuH$_2$ phase was shown to be relatively well treated within the harmonic approximation, with the predicted $T_\text{c}$ decreasing from 5-10~K to 2-5~K upon inclusion of anharmonicity. 

The good agreement between the computed $T_\text{c}$ of $P4/mmm$ PdCuH$_2$ and that measured for the hydrogenated alloy, with composition likely being closer to Pd$_{1.1}$Cu$_{0.9}$H$_{1.4}$, suggests structural similarities, on a local level. Further, the results presented herein may have broader implications for the discovery of ambient-condition superconducting hydrides, as many of the recently predicted compounds involve ternaries featuring Pd (e.g.\ Mg$_2$PdH$_6$~\cite{sanna2024prediction}), Cu (e.g.\ Li$_2$CuH$_6$~\cite{cerqueira:2024a}) and other combinations of alkaline earth and transition metal atoms~\cite{dolui:2024a}. In addition this procedure opens the door towards studies of anharmonicity using more costly electronic structure methods, such as hybrid or meta-GGA functionals, which were previously out of reach. This enables a determination of the relative stability of enthalpically close phases, which could not be separated well by the PBE functional as for example in the high pressure phases of lithium \cite{racioppi2023phase}.

To conclulde we also mention the limitations of this method. Due to the constraitn related to the Gaussian density matrix, the SSCHA is inadequate for describing molecular solids with both librational and free rotational modes \cite{doi:10.1021/acs.jctc.9b00596}. Furthermore, it fails to capture tunnelling states and struggles when the probability distribution deviates significantly from a Gaussian form \cite{monacelli2024unified}. However this technique performs well for atomic and ionic solids, where the point-like nature of the atomic or ionic building blocks suppresses large-amplitude curvilinear motion. For such systems, the SSCHA offers a significantly more computationally efficient alternative compared to Path Integral Molecular Dynamics (PIMD) \cite{doi:10.1021/acs.jctc.9b00596}.

\section{Methods}
\noindent
\subsection{Computational Details}
The CSP searches were performed using the \textsc{XtalOpt} open-source evolutionary algorithm  \cite{lonie2011xtalopt,avery2019xtalopt,XtalOpt,falls2020xtalopt}. The initial generation consisted of random symmetric structures created by \textsc{randspg}~\cite{avery2017randspg}, and duplicate structures were identified via \textsc{XtalComp}~\cite{avery2017randspg} and discarded from the breeding pool. The DFT calculations for the CSP were performed with the Vienna \emph{ab-initio} Simulation Package (VASP) version 5.4.1 \cite{Kresse:1993a, Kresse:1999a}, while the calculations of the electronic band structures, relative enthalpies, and the SSCHA were performed with the {\textsc Quantum ESPRESSO} (QE)~\cite{Giannozzi1,Giannozzi2} package (unless otherwise noted). PAW and ultrasoft pseudopotentials were used for VASP and QE, respectively, both within the Perdew-Burke-Ernzerhof (PBE)~\cite{GGA-PBE} and R$^2$SCAN meta-GGA~\cite{furness2020accurate} parametrization of the exchange-correlation functional. The Pd 4d$^{9}$5s$^{1}$5p$^{0}$4f$^{0}$, Cu 3d$^{10}$4s$^{1}$4p$^{0}$4f$^{0}$, H 1s$^{1}$2p$^0$, and Pd 5s$^1$5p$^0$4d$^9$, Cu 4s$^{1.5}$4p$^0$3d$^{9.5}$, H 1s$^1$ electronic configurations were treated explicitly for VASP and QE, respectively. The CSP relaxations for each structure were performed using a 4 step procedure where (i) the atomic positions, (ii) cell parameters and (iii+iv) cell parameters and ionic positions were relaxed. In these steps the energy cutoffs for the wavefunctions and a $\mathbf{k}$-point grid density were set to 120, 240, 300 and 400~eV, and $2\pi \times 0.1$, $2\pi \times 0.07$, $2\pi \times 0.05$, and $2\pi \times 0.025$ \AA\ $^{-1}$, respectively. The single point QE calculations were performed imposing energy cutoffs for the wavefunctions and density as 1224~eV and 12,240~eV, respectively. The integration over the Brillouin zone in the self-consistent calculations was performed with a first-order Methfessel-Paxton \cite{Methfessel} smearing of 0.272~eV coupled with a $\mathbf{k}$-point grid with density $2\pi\times 0.0175$ \AA\ $^{-1}$.

The electron-phonon coupling parameters were calculated via QE, on a $\textbf{k}$-point grid with density $2\pi \times 0.00875$ \AA\ $^{-1}$, with a Gaussian smearing of 0.1088 eV, and a $\textbf{q}$-point grid with density $2\pi \times 0.0574$ \AA\ $^{-1}$.  The superconducting critical temperature was calculated using the Allen-Dynes equation~\cite{allen1975transition} and values of the Coulomb repulsion parameter, $\mu^*$, of 0.1 and 0.15.

In the SSCHA~\cite{Lorenzo}, the displaced configurations for the various supercells used for the MTP training were generated at 300~K. The relaxation was performed by fixing the size and shape of the cell, but allowing the atoms to relax to their equilibrium positions. The MLIP package \cite{novikov2020mlip} was employed to generate an MTP of level 10, with cutoff distances of $0.9<x<9$~\AA\ . Every structure with extrapolation grade, $\gamma_\text{select}>2$, became a candidate for retraining the potential.  

With this approach, potential training was considered complete within the first 3 supercells, involving a total of 532 DFT configurations  (140 in the unit cell, 342 in the 2$\times$2$\times$2 and 50 in the 3$\times$3$\times$3 cells). The training performed at 300~K appears to be a key factor for the fast training, as temperature increases the magnitude of the atomic oscillations, thereby sampling areas of the potential that are further away from the equilibrium position. This procedure, therefore, allows for a drastic reduction in the need for further DFT calculations in SSCHA runs  performed at lower temperatures, since the smaller atomic vibrations are already contained in the area already sampled at higher temperatures. The final phonon spectra was computed at 0~K within the bubble approximation (Eq.\ 59 of Ref.\ \cite{Lorenzo}). 

The anharmonic $T_\mathrm{c}$s were calculated using the phonon eigenvalues and eigenvectors obtained through the SSCHA relaxation, along with the electron-phonon coupling matrix elements obtained from the classical DFPT calculations for the same crystal structure, as the SSCHA relaxation did not result in any significant structural changes.\\

\noindent{\bf Code Availability.}
The code will be released in a subsequent publication, and in the meanwhile, may be obtained from the corresponding authors upon reasonable request. \\

\noindent\textbf{Data Availability}
The datasets generated during and/or analyzed during the current study are summarized in the supplementary information, and are available from the corresponding author on reasonable request. \\

\noindent\textbf{Acknowledgements}
Funding for this research is provided by the National Science Foundation, under award DMR-2136038. Calculations were performed at the Center for Computational Research at SUNY Buffalo (http://hdl.handle.net/10477/79221).  \\

\noindent\textbf{Author Contributions}
The project was conceived and supervised E.Z. F.B.\ developed the method and performed the calculations and analysis. F.B.\ wrote the manuscript, which was edited by E.Z.  \\

\noindent\textbf{Competing Interests:} The authors declare no competing interests. \\


\bibliography{bibliography}
\end{document}


\maketitle

\clearpage
\newpage

\tableofcontents

\clearpage
\newpage

\section{Structural Parameters}
\begin{table}[h!]
	\centering
\caption{\textbf{Structural parameters.} The table reports the structural parameters for the phases obtained through the evolutionary structure search, after full structural relaxation.}

\begin{tabular}{|c|c|c|}
\hline
 Composition (Spacegroup) & Lattice Parameters & Wyckoff positions  \\
 \hline \hline
                    &  $a$ = 2.950 ({\AA})  &  Pd \textbf{1a}  0.0000 0.0000 0.0000\\
 PdCu ($Pm\bar{3}m$)  &                    &  Cu \textbf{1b}  0.5000 0.5000 0.5000\\
                    &                    &                                      \\
 \hline
                    & $a$ = 2.770 ({\AA}) &  Pd \textbf{1a}  0.0000 0.0000 0.0000\\
 PdCuH ($P4/mmm$)       & $c$ = 3.695 ({\AA}) &  Cu \textbf{1d}  0.5000 0.5000 0.5000\\
                    &                  &  H  \textbf{1c}  0.5000 0.5000 0.0000\\
 \hline
                    & $a$ = 2.853 ({\AA}) &  Pd \textbf{1a}  0.0000 0.0000 0.0000\\
  PdCuH$_2$ ($P4/mmm$)  & $c$ = 3.751 ({\AA}) &  Cu \textbf{1d}  0.5000 0.5000 0.5000\\
                    &                  &  H  \textbf{1b}  0.0000 0.0000 0.5000\\
                    &                  &  H  \textbf{1c}  0.5000 0.5000 0.0000\\  
\hline
                      & $a$ = 3.825 ({\AA}) &  Pd \textbf{4a}  0.0000 0.0000 0.0000\\
PdCuH$_2$ ($I4_1/amd$)  & $c$ = 8.440 ({\AA}) &  Cu \textbf{4b}  0.5000 0.5000 0.0000\\
                      &                   &  H  \textbf{8e}  0.0000 0.0000 0.7665\\ 
                      \hline
\end{tabular}
\end{table}

\newpage
\section{Supplementary Note 1: Accuracy of the Moment Tensor Potential}

To evaluate the accuracy of the developed potential the root mean square (RMS) prediction errors for energies, forces, and stresses (EFS) were analyzed for the $P4/mmm$ PdCuH$_2$ structure as a function of the supercell size, as shown in Supplementary Figure \ref{Fig:POTENTIAL_CONVERGENCE}. The test set of structures was generated using the SSCHA. This set contained 200 configurations for the 1$\times$1$\times$1, 200 configurations for the 2$\times$2$\times$2, 200 configurations for the 3$\times$3$\times$3 supercells, and 70 configurations for the 4$\times$4$\times$4 supercell. This test set of structures was distinct from the training set. We employed geometries generated by the SSCHA for testing, since these would be representative of structures encountered during the SSCHA+MLIP procedure.
The MTP was trained, as described in the main text, for the 1$\times$1$\times$1 unit cell. After training was complete, it was employed to predict the EFS for the aforementioned test set of configurations generated for the 1$\times$1$\times$1, 2$\times$2$\times$2, 3$\times$3$\times$3 and 4$\times$4$\times$4 cells. As seen in the figure (dark green line), while this potential did a relatively good job of predicting the EFS for the unit cell, the errors in the energy for larger cells were as large as 200 meV/atom (for the 3$\times$3$\times$3). It performed similarly poorly for predicting the forces and stresses of the larger cells. The MTP trained on the 2$\times$2$\times$2 cell (light green line) showed a dramatic improvement with the RMS errors for the energies being less than 3~meV/atom for all of the unit cell sizes tested. However, the force errors for the 3$\times$3$\times$3 were quite large. Training on the 3$\times$3$\times$3 cell (orange line) was required to reduce the RMS prediction errors for the EFS on the 4$\times$4$\times$4 cell to $<$ 0.5~meV/atom, $<$ 50~meV/\AA\  and $<$ 0.5~GPa, respectively. While the RMS error for the energies and stresses for the unit cell calculated with this potential were larger, this is not concerning because in the SSCHA+MLIP procedure MTPs trained on larger cells are never used for predictions on smaller ones.

\begin{figure*}[h]
\centering
\includegraphics[scale=0.5]{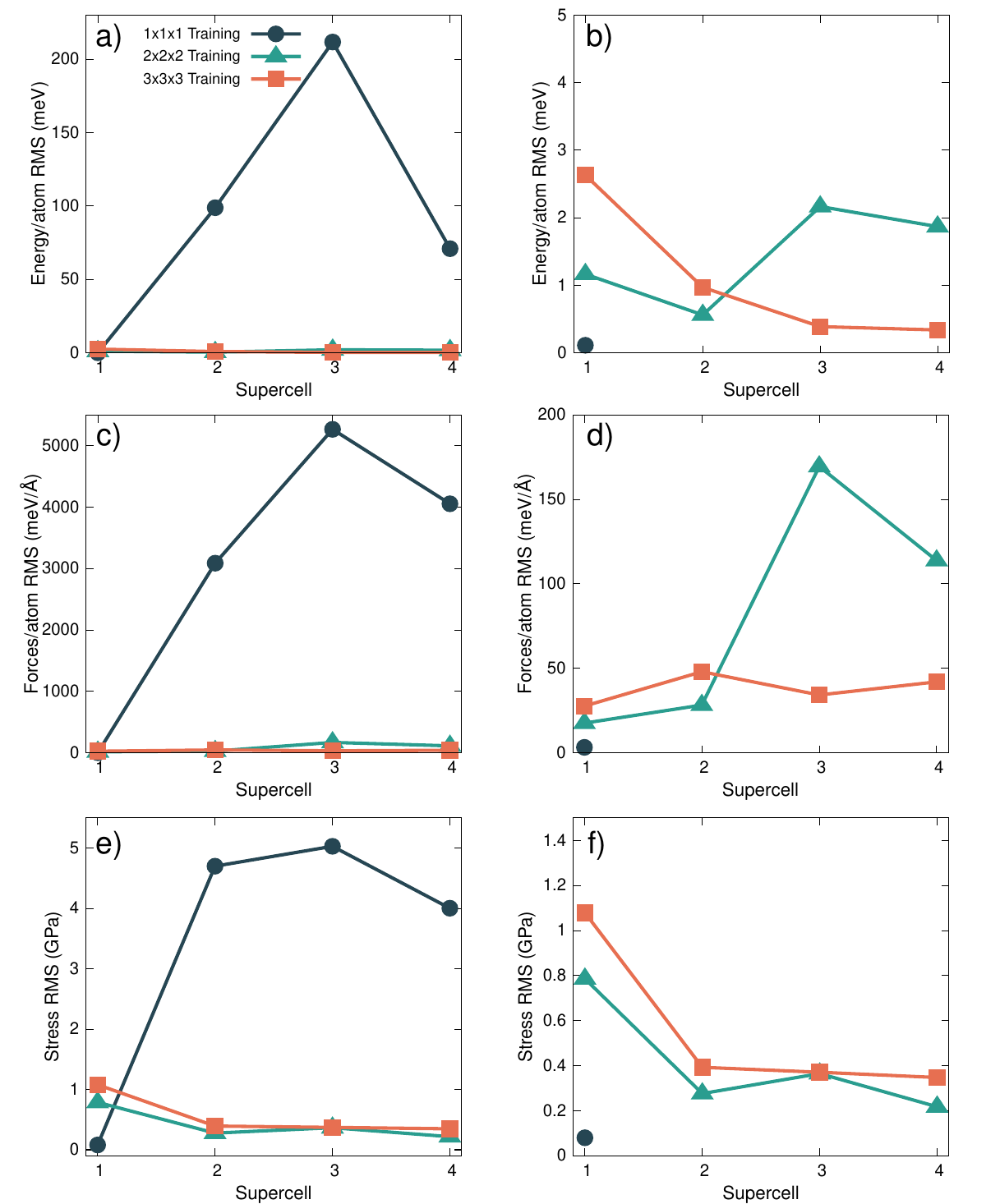}
\caption{\label{Fig:POTENTIAL_CONVERGENCE} \textbf{Accuracy of the potential.} The figure illustrates the root mean squared (RMS) prediction errors for energies, forces, and stresses as a function of the supercell size for the $P4/mmm$ PdCuH$_2$ structure. Panels (a) and (b) show errors for energies, panels (c) and (d) for forces, and panels (e) and (f) for stresses. The dark green lines represent results from training on the unit cell alone, the light green lines include training on both the unit cell and the 2$\times$2$\times$2 supercell, and the orange lines incorporate training on the 3$\times$3$\times$3 and smaller cells. The right-hand panels provide a closer view of the data from the left-hand panels, with lines included as a visual guide. }
\end{figure*}

\clearpage

\begin{figure}
    \centering
    \includegraphics[width=0.7\textwidth]{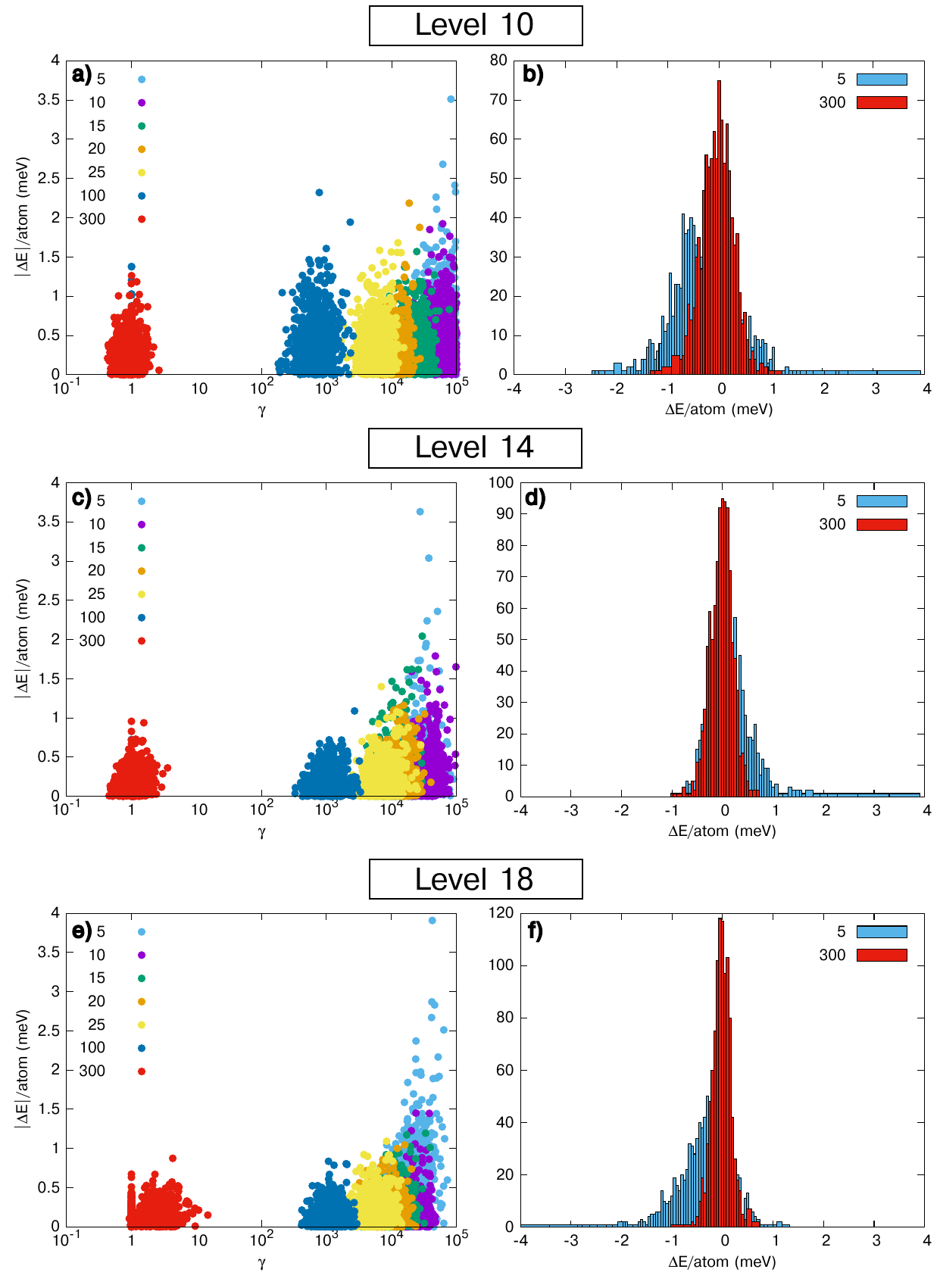}
\caption{\textbf{True error}. The figure shows the true error ($\Delta E$) and the extrapolation grade $\gamma$  (a, c, e), and the histogram of the true error distribution (b, d, f) for different training set sizes (legend) and levels of MTP for $P4/mmm$ PdCuH$_2$. The dataset consists of 1000 structures generated using the SSCHA on a 2$\times$2$\times$2 supercell with 32 atoms. With a greater amount structures in the training set both the values of $\gamma$ and $\Delta E$ are reduced. When the number of structures comprising the training set was large enough so that all $\gamma$ values fell below 300, there did not appear to be a relevant decrease in the true error when the number of structures in the training set was increased.}
\label{fig:gamma}
\end{figure}

Furthermore, we investigated how the true error, $\Delta E$, and the extrapolation grades,  $\gamma$, of the set of structures outside the training set encountered during the SSCHA were affected by the size of the training set (Figure \ref{fig:gamma}). The calculations employed a 2$\times$2$\times$2 supercell and 1000 configurations for the noted MTP levels. The analysis reveals that increasing the number of structures used for training decreases both the $\Delta E$ as well as $\gamma$. Furthermore, the choice of the MTP level had the most pronounced impact on $\Delta E$, irrespective of how many structures were used for training. The structures with a $\gamma<$10 all possessed $\Delta E<$1.5~meV/atom, irrespective of the MTP level, suggesting that the extrapolation was reliable, as suggested by Novikov and co-workers (\url{https://iopscience.iop.org/article/10.1088/2632-2153/abc9fe}).

\begin{figure}
    \centering
    \includegraphics[width=0.8\textwidth]{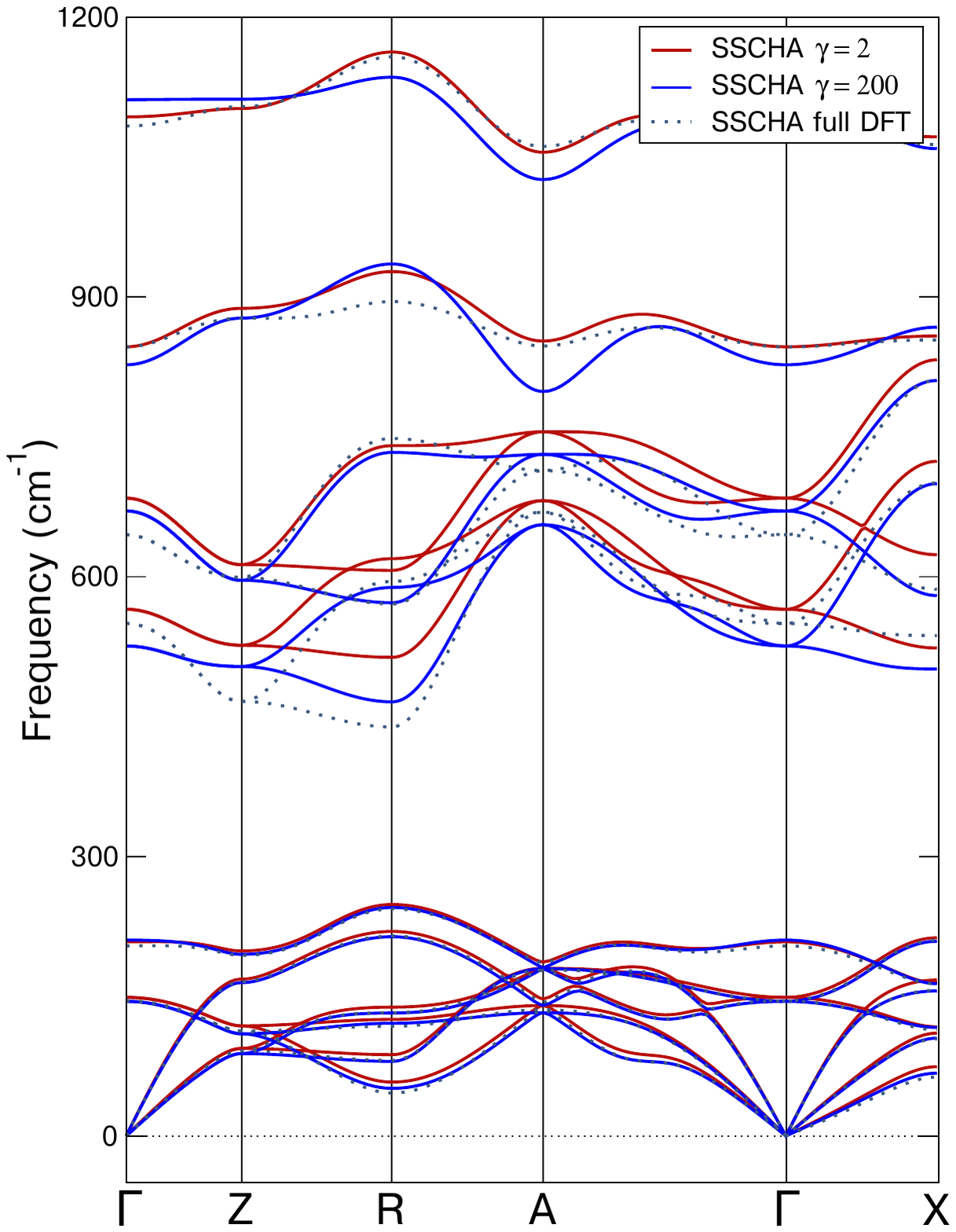}
\caption{\textbf{Phonon band structures}. The figure shows the anharmonic phonon spectra calculated for a 2$\times$2$\times$2 supercell with different values of the extrapolation grade $\gamma$ and an MTP level of 10 for the $P4/mmm$ PdCuH$_2$ structure. The dotted lines correspond to the full DFT calculations while the red and blue lines corresponds to values of $\gamma$ of 2 and 200, respectively. Already for $\gamma=200$ the phonon spectra is reasonably reproduced. An improvement is obtained for the highest phonon frequencies using $\gamma=2$, producing a more accurate prediction of the higher optical brances. For additional accuracy a higher level of MTP is likely necessary.}
\label{fig:Gamma_Phonons}
\end{figure}

Finally, we investigated the accuracy of the SSCHA phonon band structures obtained for $\gamma$ values of 2 and 200 against the full DFT results in Figure \ref{fig:Gamma_Phonons} using an MTP level of 10. The phonon bands obtained with $\gamma=2$ provide significantly better agreement with the DFT results for the higher optical branches, but higher MTP levels might be needed to improve the accuracy of the intermediate optical branches.

%

\newpage
\section{PdCuD$_2$ Phonons}
\begin{figure*}[h]
\centering
\includegraphics[scale=0.6]{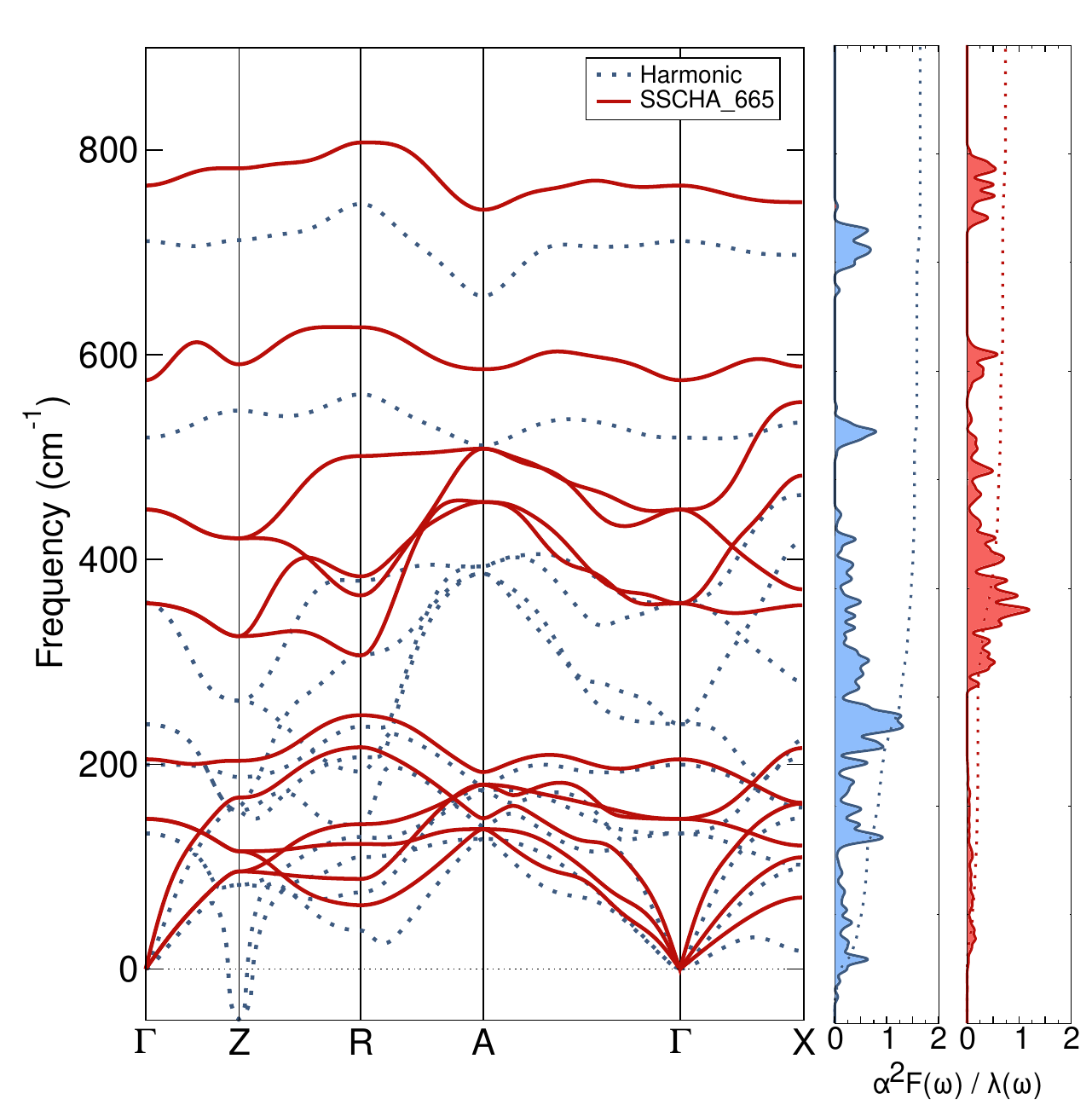}
\caption{\label{Fig:PdCuD} \textbf{PdCuD$_2$ Phonons.} The phonon band structure, Eliashberg spectral function, $\alpha^2F(\omega)$, and the integral of the electron phonon coupling, $\lambda(\omega)$,  for  $P4/mmm$ PdCuD$_2$ calculated with classical nuclei (blue dashed lines and shading) and incorporating the effects of anharmonicity and quantum nuclei (red solid lines and shading).}
\end{figure*}
\newpage
\section{Computational Resources}
\begin{table}[h!]
	\centering
\caption{\textbf{Computational resources.} The table reports the number of configurations required and the number of configurations effectively calculated through DFT in the active training of the potential up to the 3$\times$3$\times$3 supercell  for the $P4/mmm$ PdCuH$_2$ structure. We also report the time in hours for a single DFT calculation (DFT time), the effective number of hours required by the training (Total time) against the time actually spent (Actual time). }

\begin{tabular}{|c|c|c|c|c|c|}

\hline \
  Supercell & Total Conf & DFT Conf & DFT time (h) & Total time (h) & Actual time (h)   \\
 \hline \hline

           1$\times$1$\times$1 & 140  & 140 & 1.6  & 224    & 224  \\
           2$\times$2$\times$2 & 1500 & 342 & 12   & 18,000  & 4104 \\
           3$\times$3$\times$3 & 2180 & 50  & 72   & 156,960 & 360  \\
\hline \hline
                              &      &     &      & 175,184 & 4688 \\                    
\hline 
                              
\end{tabular}
\end{table}

\section{Convergence Tests}
\begin{figure*}[h]
\centering
\includegraphics[width=\textwidth]{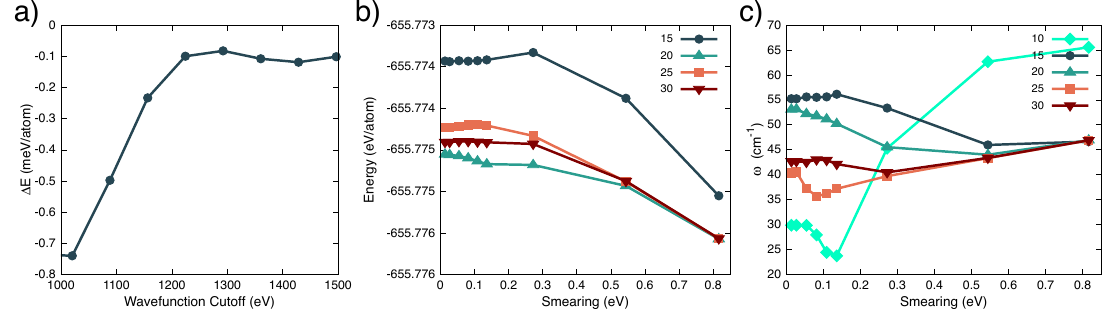}
\caption{\label{Fig:ConvTest} \textbf{Convergence tests.} The panels in the figure report the convergence tests  for the $P4/mmm$ PdCuH$_2$ structure performed for determining the wavefunction cutoff using a 20$\times$20$\times$16 $k$-point mesh and 0.032~eV smearing (a), and the convergence of the energy (b) and the lowest acoustic phonon frequency at $\Gamma$ (c) as a function of smearing and a 10$\times$10$\times$8 (aqua green), 15$\times$15$\times$12 (dark green), 20$\times$20$\times$16 (green), 25$\times$25$\times$20 (orange) and 30$\times$30$\times$24 (red) $k$-point mesh. }
\end{figure*}
\clearpage

\section{Choice of Density Functional}
\begin{figure*}[h]
\centering
\includegraphics[width=0.8\textwidth]{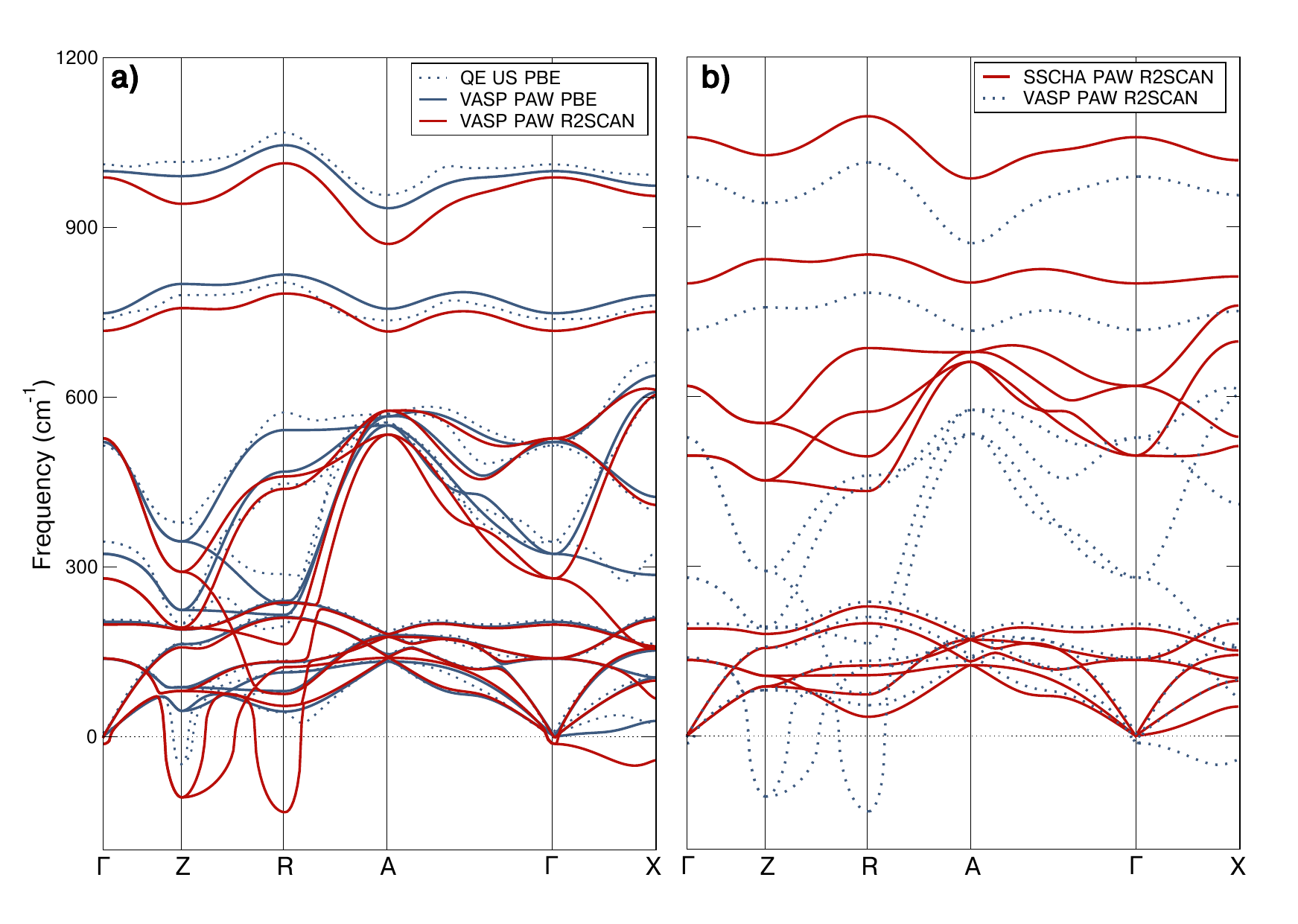}
\caption{\label{Fig:ConvTest} \textbf{Functional tests.} Harmonic phonon spectra with different pseudopotentials and functionals for the $P4/mmm$ PdCuH$_2$ structure (a). The dotted lines correspond to the calculations performed with Quantum Espresso (QE) using ultrasoft pseudopotentials with the PBE exchange-correlation functional. The blue and red lines correspond to the calculations performed with the Vienna \emph{Ab initio} Simulation Package (VASP) using the Projector Augmented Wave (PAW) method with the PBE (blue) and R2SCAN (red) exchange-correlation functionals. The phonon spectra resemble each other, exhibiting the same type of soft modes. The right panel reports the classical and quantum anharmonic phonon spectra calculated using the PAW method with the R2SCAN functional (b). The dotted lines correspond to the classical results while the red lines correspond to the SSCHA quantum anharmonic phonons.}
\end{figure*}

\clearpage
\begin{figure*}[h]
\centering
\includegraphics[width=0.8\textwidth]{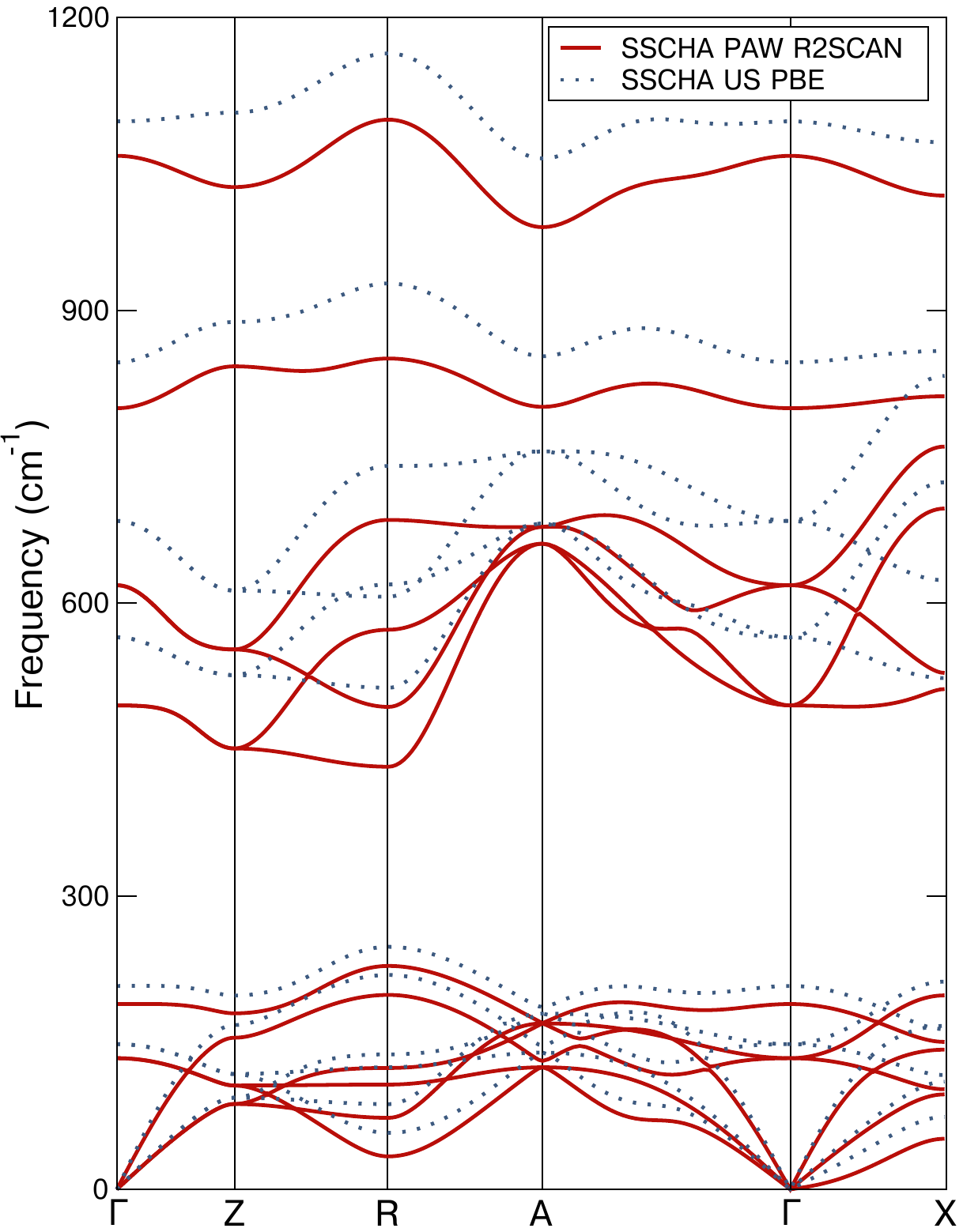}
\caption{\label{Fig:ConvTest} \textbf{SSCHA Functional tests.} SSCHA phonon spectra obtained with ultrasoft pseudopotentials using the PBE exchange correlation functional (blue dotted lines), and with the PAW method and the R2SCAN exchange correlation functional (red lines) for the $P4/mmm$ PdCuH$_2$ structure.
The R2SCAN functional introduces a red shift of the phonons, but produces the same qualitative results as PBE.}
\end{figure*}